\documentclass[11pt]{article}

\usepackage{authblk}        % for affiliations
\usepackage{orcidlink}      % for ORCID icons
\usepackage{hyperref}       % for clickable links (including ORCID)

\usepackage{amsmath}
\usepackage{amsfonts}
\usepackage{dsfont}
\usepackage{url}
\usepackage{doi}

\usepackage{epigraph}

\newcommand{\Sec}[1]{Sec.~\ref{#1}}

\usepackage[utf8]{inputenc}
\usepackage[T1]{fontenc}

\usepackage[margin=1in]{geometry}

\title{\bf Martin Davis: An Overview of his Work in Logic, Computer Science, and Philosophy}

\author[1]{Liesbeth De Mol\,\orcidlink{0000-0001-8520-2493}}
\author[2]{Yuri V. Matiyasevich\,\orcidlink{0000-0001-7046-3746}}
\author[3]{Eugenio G. Omodeo\,\orcidlink{0000-0003-3917-1942}}
\author[4]{Alberto Policriti\,\orcidlink{0000-0001-8502-5896}}
\author[5]{Wilfried Sieg\,\orcidlink{0000-0002-7130-0524}}
\author[6]{Elaine J. Weyuker,\orcidlink{0000-0002-1660-199X}}

\affil[1]{UMR 8163 Savoirs, Textes, Langage, CNRS - Universit\'e de Lille, \newline Rue du Barreau, 59653 Villeneuve d'Ascq, France}

\affil[2]{Laboratory of mathematical logic and discrete mathematics, \newline St.\ Petersburg Department of Steklov Mathematical Institute of the Russian Academy of Sciences, 27, Fontanka River embankment, 191023 St.\ Petersburg, Russia}

\affil[3]{Department of Mathematics, Informatics, and Geosciences, \newline University of Trieste, via Valerio 12/1, 34127 Trieste, Italy}

\affil[4]{Department of Mathematics, Computer Science, and Physics, \newline University of Udine, via delle Scienze 206, 33100 Udine, Italy}

\affil[5]{Department of Philosophy, Carnegie Mellon University, \newline 5000 Forbes Avenue, 15213, PA, USA}

\affil[6]{Department of Computer Science, University of Central Florida, \newline 4000 Central Florida Blvd, Orlando,  32816, FL, USA}
\date{}  % No date

\begin{document}

\maketitle

\begin{abstract}
In his autobiographic essay written in 1999, ``From logic to computer science and back'', Martin David Davis (3/8/1928--1/1/2023) indicated that he viewed himself as a logician \emph{and} a computer scientist. He expanded the essay in 2016 and expressed a new perspective through a changed title, ``My life as a logician''. He points out that logic was the unifying theme underlying his scientific career. Our paper attempts to provide a consistent vision that illuminates Davis' successive contributions leading to his landmark writings on computability, unsolvable problems, automated reasoning, as well as the history and philosophy of computing.
\end{abstract}

\epigraph{When I was invited to
speak at the weekly colloquium of the Berkeley mathematics department, I took the opportunity to express my view of computability theory as an autonomous branch of mathematics. Tarski, who was in the audience, took strong exception during the discussion after my talk.}{\textit{Martin~Davis in \cite{Davis20}}}

\section{Personal and academic life.}
Martin David Davis was born in 1928 to immigrant Polish Jewish parents in the Bronx, New York City, which was then a largely working class community. A few years before Davis started high school, New York opened a specialized public high school, The Bronx High School of Science, offering a highly enriched free education with admission determined by a student's exam score.  Davis graduated in 1944 and then attended the City College of New York, which also provided a free education to New York City students. He majored in mathematics and studied under Emil Leon Post, a man whom Davis greatly admired and who highly influenced Davis's life work. After graduation in 1948, Davis went to Princeton to study for his Ph.D. in mathematical logic. His dissertation advisor was Alonzo Church, another major figure in the early world of computation theory. He earned his degree in 1950.

Davis spent the decade following his doctorate, as he recalls in \cite{Dav16} (see also \cite{CHOPS24}), moving among various academic institutions: University of Illinois at Urbana-Champaign, Institute for Advanced Study in Princeton, University of California--Davis, Ohio State University, Hartford Graduate Division of Rensselaer Polytechnic Institute. In 1951 Davis married Virginia Palmer\footnote{Their marriage would last 71 years till they both passed away on January 1, 2023.} with whom he had two sons: Harold and Nathan. Around 1960, Davis settled in New York City and developed a program in logic at Yeshiva University. From 1965 to his retirement in 1996, he was a faculty member at New York University's Courant Institute of Mathematical Sciences, first as a professor of Mathematics and later of Computer Science. He served as Chair of Computer Science from 1988 to 1990. After retiring from NYU, he lived in Berkeley and continued to work as a ``Visiting Scholar'' at the university `where Alfred Tarski had developed a world-class logic group' \cite[p.~{31}]{Dav16}.

\section{Computability and unsolvability.}\label{Sec:computabilityAndUnsolvability} 
The main focus, henceforth, will be on aspects of Davis's activity that have moved forward computer science and its foundations.
Unavoidably, his contributions to logic as such (for example in \cite{Davis-Fechter-91}) and his explorations of the foundations of mathematics and analysis will remain in the shadows, except for a quick account in \Sec{sec:nonstandardAnalysis}.

\smallskip

In May of 1950, Davis's dissertation \emph{On the Theory of Recursive Unsolvability} was recommended for acceptance by Princeton's Department of
Mathematics.\footnote{ Davis had just turned twenty-two. Recursive functions are a key formalism within computability theory. Essentially, these are functions that work on natural numbers. They were used as a formal model for effective calculability by Church.}
%LDM Recursive functions are a key formalism within computability theory. Essentially, these are functions that work on natural numbers. They were used as a formal model for effective calculability by Church, resulting in Church's thesis. Note that Church also referred to the $\lambda$-definable functions here. These were proven to be logically equivalent to general recursive functions.   
When expressing his gratitude
``toward the various persons whose help has meant so much in the preparation of this
dissertation'', Davis thanked Alonzo Church for reading the manuscript
several times and offering ``valuable criticisms and suggestions''. That standard ``Thank you!'' of a doctoral student to his advisor is followed by a
striking acknowledgement of intellectual debt: ``It was Professor Emil L. Post who first
introduced me to the theory of recursive functions, and who is largely responsible for
my present point of view toward the subject.'' That point of view was not just Davis's
point of view in 1950, but persisted throughout his career.
%LDM: perhaps we can condense this paragraph? EG: While Davis showed himself grateful to his advisor Church, it is clear from the thesis already that it was Emil L. Post to whom he was mostly indebted. 

The thesis---as we will argue next---is a remarkable document.\footnote{In his NAMS Interview \cite[p.~563]{Jac08}, Davis asserted: ``By the time I graduated [from] City College, I knew I wanted to be a logician. I had written a term paper for an advanced logic course in the Philosophy Department, which in a way was a first draft of part of what was later my dissertation.''} In Part I, Kleene's theory of recursive functions is recast, generalized, and shown to be equivalent to a similar one of Post's.
One deep expression of the equivalence is the theorem stating that a set is
recursively enumerable if and only if it is normal.  That is, any set of natural numbers that is the range of a recursive function can be generated by a normal system and vice versa.  Normal systems were introduced by Post and are essentially very simple rewrite systems for strings.
In Part II, Kleene's arithmetical
hierarchy is extended into the transfinite along the constructive second number class
generating the hyperarithmetical hierarchy. Central results were presented in a talk at the 1950 International Congress of Mathematicians; see \cite{Davis52}. They, together with then recent results, are also described in the last sections of \cite{Dav58}. A detailed development of the theory of hyperarithmetical sets and Davis's seminal role is found in Moschovakis's \cite{Moschovakis16}. Finally, Part III focuses on an ``irresistible''
problem, namely, Hilbert's Tenth Problem. The thesis exhibits characteristic traits of
Davis's work; namely, a beautifully clear mathematical exposition as well as a deep
desire to show the broad context for the introduction of mathematical notions and for their systematic development. 

The broad context is sketched in the Introduction to the thesis. There, Davis not only traces the development of recursive function theory but also describes its ``most spectacular results''. These include a number of
logically equivalent results related to the impossibility of solving
certain mathematico-logical problems in a finite number of steps. The
halting problem for Turing machines is one such instance. Most important
to Davis' exposition then was Post's elegant mathematical formulation of
Turing machines to offer an alternative proof to the unsolvability of
the halting problem which was then used to prove that the word problem
for semi-groups is also unsolvable. This then recent
research result was presented in Davis's 1958 textbook \emph{Computability and Unsolvability}, \cite{Dav58}; a
broad swath of unsolvable problems is found in Davis's \cite{Davis77a}. Not only is Post's
result established in the textbook, but Post's way of defining Turing machines, as production systems, provided the
basis for the exposition of computation theory. Indeed, central results coming from mathematical logic (incompleteness and undecidability theorems) are established on that very basis.
It has to be emphasized most strongly that Davis in 1958 thought of the subject no longer as ``recursive function theory'' (as he had done in his dissertation) but rather as ``computability theory''. At the time, \emph{Computability and Unsolvability} was one of the founding texts of an emerging new field, computer science. 

This deep connection to theoretical computer science is seen easily from a contemporary perspective; in 1983 it is directly pointed out in the textbook co-authored with Elaine Weyuker, \emph{Computability, Complexity, and Languages --- Fundamentals of Theoretical Computer Science} \cite{DavWey83}.\footnote{A much enriched 2nd edition, \cite{DavSigWey94}, co-authored with Ron Sigal appeared in 1994.} In the Preface they write: ``Theoretical computer science is the mathematical study of models of computation. As such, it originated in the 1930s, well before the existence of modern computers, in the work of the logicians Church, G\"odel, Kleene, Post, and Turing. This early work has had a profound influence on the practical and theoretical developments of computer science. Not only has the Turing-machine model proved basic for theory, but the work of these pioneers presaged many aspects of computational practice that are now commonplace.''

%addition from Wilfried: "It should be noted that Davis, in 1958, thought of hte subject no longer as "recursive function thery" but rather as "computability theory"; that change in conceptual perspective was most stringly urged by other logicians only in the 1990s. For Davis this quite dramatic change was based on the fact that Post"
%LDM my proposal to continue as follows: completely captured Turing's way of analyzing mechanical calculations of a human being via 'production systems" which was picked up later y Turing in \cite{Turing50}.  The deep underlying conceptual unity of Post's and Turing's approaches is palpable in Martin's book \textit{Computability and unsolvability}. It is explicit .... 

For Davis this quite dramatic change of conceptual perspective from recursive function theory to computability theory was based on the fact that Post completely captured Turing's way of analyzing the mechanical calculations
of a human being via ``production systems'' given in \cite{Turing36}. Indeed, in \cite{Turing50}, Turing extended Post's techniques to establish the undecidability of the word
problem for semigroups with cancellation; then in \cite{Turing54}, Turing used Post's concept of
deterministic production systems (and dubbed them ``substitution puzzles'') for two
goals: (1) to discuss the methodological problem of articulating a mathematically
precise and adequate concept of computation and (2) to prove central undecidability
results. The deep underlying conceptual unity of Post's and Turing's approaches is
palpable in Davis's book. It is explicit in a paper Davis and Sieg published in 2016 \cite{DavSieg16},
``Conceptual Confluence in 1936: Post and Turing''. This is standardly discussed under
the headings of Church's thesis and Turing's thesis.\footnote{\label{footnote:ChurchThesis}Church's thesis states that any effectively calculable function is general recursive (and $\lambda$-definable). It is
extensionally equivalent to Turing's thesis stating that any computable function is computable by a Turing machine.
At the core of the theses is the methodological question of the legitimacy of identifying an informal notion (effective
calculability, computability) with a rigorous mathematical concept (general recursiveness, Turing machine
computability).}

In his dissertation Davis had listed ``spectacular results'' of, what he called then, recursive function theory. He made a real contribution to this field by editing a collection of most of the original papers in which these results had been established. The collection was published as \emph{The Undecidable: Basic papers on undecidable propositions, unsolvable problems and computable functions}. To this day, it is a fundamental source book for anyone studying the history, philosophy and theory of computability. Post has a major place in the volume as it includes not only some of his published papers but also the fascinating and previously unpublished \emph{Absolutely unsolvable problems and relatively undecidable propositions --- account of an anticipation}. That paper contains Post's account of how he anticipated results of Church, G\"odel and Turing in the early 1920s.\footnote{Post submitted this paper to the American Journal of Mathematics in 1941. However, the editor-in-chief Hermann Weyl rejected it, saying that the \textit{Journal} was ``\textit{no place for historical accounts}'' (quoted in \cite{davis+emil+1994}). At the same time, Weyl suggested focusing on the central mathematical results; that led to a shortened version that was ultimately published as \cite{post+formal+1943}. It contains Post's normal form theorem and a statement of Post's first thesis.}  
Of course, in \emph{The Collected Works of Emil L. Post}, all of Post's mathematical papers were published. Davis finished editing the volume in 1993 and remarked in the Preface that many of Post's students were deeply influenced by him and pursued successful careers in mathematics. About himself Davis wrote: ``I am the only one whose own research interests have followed in the directions he pioneered, and so I feel it particularly appropriate that I play the role of his editor.''

Davis worked on the history of computing not only through editing the two volumes described above, but by original explorations. Since he contributed to laying out a theory of computation, it should not come as a surprise that he also contributed to philosophical aspects of the field. We return to this in \Sec{Sec:HistoryAndPhilosophy}. However, there is one particular mathematical challenge that compelled his passionate engagement starting with his thesis, namely, Hilbert's Tenth Problem. The absolutely fascinating developments related to this problem are the topic of the next section.

 %\color{blue}
\section{Hilbert's Tenth Problem.} %?% \label{Sec:H10}

One cannot overestimate the role played by Davis
in resolving  Hilbert's Tenth Problem. It is one of the 23 mathematical problems posed by David Hilbert \cite{Hilbert00,Hilbert0102} in 1900 at the Second International Congress of Mathematicians held in Paris.

Among these 23  problems the tenth is the only one
which
can be considered today as belonging to  Computer Science
(which did not exist in 1900). In this problem Hilbert
asked about the solvability of \emph{Diophantine equations}.
They are equations of the form
\begin{equation}\label{dioph}
 P(x_1,\dots,x_m)=0
\end{equation}
where $P$ is a polynomial with integer coefficients.

The equations are named after the Greek mathematician
Diophantus of Alexandria who lived in the 3rd century  AD.
Polynomial equations had been considered by Greek mathematicians before him but in a geometrical manner. For example,
to solve equation
\begin{equation}\label{x2}
  x^2=2
\end{equation}
one had to draw a square with the unit side and
take its diagonal.

The novelty introduced by Diophantus was as follows:
he was seeking solutions of polynomial equations
in (positive) rational numbers. For him equation
\eqref{x2} had no solution.

Hilbert in his Tenth Problem asked about solving
Diophantine equations in integers.
After Diophantus's time  mathematicians found solutions of
many Diophantine equations, and established that many others
had no solutions. This was done by diverse, sometimes
\emph{ad hoc}, methods (a well-known example is
Fermat's Last Theorem). Hilbert asked  to find
a \emph{universal method} applicable to every Diophantine equation.

In today's terminology we would say that Hilbert asked to
find an algorithm. However, he did not use this word ---
at that time there was no mathematically rigorous
general notion of algorithm. It was developed only
in the 1930s by Kurt G{\"o}del, Alan Turing, Emil Post,
Alonzo Church and other logicians.
 They gave
several equivalent definitions via different models of computation such as \emph{Turing machines},
\emph{$\lambda$-calculus},
\emph{recursive functions} etc.

Davis's interest in Hilbert's Tenth Problem 
 was inspired by his teacher Emil
Post who wrote in his paper \cite{Post44} that this problem ``begs for an unsolvability proof''.

Davis quickly devised a conjecture, first announced in \cite{Davis50a},
which
would imply the non-existence of the required algorithm.
However, the statement of his conjecture did not contain
the word ``algorithm'' either. Dealing with Diophantine equations, it is more convenient to  work with another concept closely related to algorithms.

A set $\mathfrak M$ of  $n$-tuples of natural numbers is called \emph{listable}
(aka \emph{semidecidable, recursively enumerable}) if all its elements
(and only those) can be printed by some algorithm (computer program which may work for an arbitrary finite number of steps). A set $\mathfrak M$ of  $n$-tuples of natural numbers is called \emph{decidable},  if 
there exists an algorithm which for given arbitrary 
$n$-tuples of natural numbers tells us after a finite number of steps whether this  $n$-tuples belongs to
$\mathfrak M$ or not. The existence of undecidable listable sets was a major discovery in computability theory.

A set $\mathfrak M$ of  $n$-tuples of natural numbers is called \emph{Diophantine} if
there exists a polynomial $P(a_1,\dots,a_n,x_1,\dots,x_m)$ with integer coefficients
such that
\begin{equation}\label{dio}
  \langle a_1,\dots,a_n \rangle  \in \mathfrak M \iff
  \exists x_1,\dots,x_m [P(a_1,\dots,a_n,x_1,\dots,x_m)=0].
\end{equation}
An equivalence of this form is called a \emph{Diophantine representation}
of the set~$\mathfrak M$.

It is easy to see that every Diophantine set is listable. Martin Davis conjectured that
the inverse is also  true.

\

{\bf Martin Davis conjecture.} \emph{Every listable set is Diophantine.}

\

This conjecture implies the negative solution of
Hilbert's Tenth Problem. Indeed, it suffices to 
take 
an undeciable listable set for $\mathfrak M$ 
in \eqref{dio}; 
then no algorithm would decide whether the 
Diophantine equation
on the right-hand side of \eqref{dio} 
has a solution in $x_1,\dots,x_m$  for given values 
of the 
parameters $a_1,\dots,a_n$.

Davis's conjecture is much stronger than what would be sufficient for proving the undecidability of Hilbert's Tenth Problem. Namely, it would be enough  to allow $P$ in the definition \eqref{dio} to be any function which becomes a polynomial in $x_1,\dots,x_m$ after substituting numerical values for $a_1,\dots,a_n$.
As an example of such a generalization, one could allow
the parameters (but not the unknowns)  to occur 
in the exponents. Such a weaker form of Davis's approach 
to Hilbert's Tenth Problem 
was suggested by Anatolii Maltsev \cite{Maltsev} (see also
\cite[Comments to Ch. 5]{Matiyasevich93a}).

Matiyasevich once called Davis's conjecture ``bold''.
Indeed, he had few informal arguments supporting it.
On the other hand, the conjecture had, in addition to the undecidability of Hilbert's Tenth Problem, many corollaries, some of which were quite striking. For example, the conjecture implies that the (evidently listable) set of all prime numbers is exactly the set of all non-negative values assumed by a certain polynomial for all possible integer values of its variables.

The theory of computability can be developed either in
terms of algorithm or in a dual form via listable sets.
The second approach was used by Davis's teacher Emil Post
\cite{Post43} and later 
by G.\,S.\,Tseitin  \cite{Tseitin64} and P.\,Martin-L\"of
 \cite{MartinLof}.

In 1976 L.\,Adleman and K.\,Manders \cite{AdlemanMa76}  expressed  the relationship
between the two concepts by introducing the notion of
\emph{non-deterministic Diophantine machines}, NDDMs for short.
Such a machine is specified by a parametric Diophantine equation
%\eqref{arbit}
and works as follows: on input
$\langle a_1,\dots,a_n\rangle$ it guesses the numbers
$x_1,\dots, x_m$ and calculates the value of the polynomial; if
it vanishes,
then the $n$-tuple
$\langle a_1,\dots,a_n\rangle$ is accepted
(is recognized as belonging to the set).

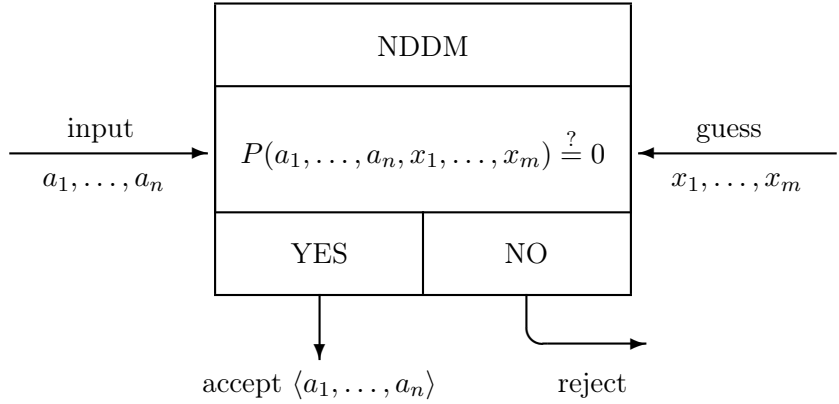
\setlength{\unitlength}{1.1mm}
\thicklines
\begin{center}
\begin{picture}(100,55)(0,0)
\put(25,15){\framebox(50,35){}}
\put(25,40){\makebox(50,10){NDDM}}
\put(25,25){\framebox(50,15){$P(a_1,\dots,a_n,x_1,\dots,x_m)
\stackrel{?}{=}0$}}
\put(100,32){\vector(-1,0){24}}
\put(0,32){\vector(1,0){24}}
\put(50,15){\line(0,1){10}}
\put(37.5,15){\vector(0,-1){8}}
\put(65,15){\oval(5,12)[bl]}
\put(65,9){\vector(1,0){12}}
\put(7,34){input }
\put(4,28){$a_1,\dots,a_n$}
\put(83,34){guess }
\put(80,28){$x_1,\dots,x_m$}
\put(25,15){\makebox(25,10){YES}}
\put(50,15){\makebox(25,10){NO}}
\put(25,-01){\makebox(25,10){accept
$\langle a_1,\dots,a_n\rangle$}}
\put(58,-1){\makebox(25,10){reject}}
\end{picture}
\end{center}

Davis's conjecture can be restated as follows.

\

\textbf{Davis's conjecture}. \emph{Non-deterministic Diophantine
machines are as powerful as, say, Turing machines}.

\

The rationale behind the introduction of yet another
model of computations was as follows.
Diophantine machines are essentially non-deterministic,
and have explicit  separation between the non-deterministic
part  (guessing the values of $x_1,\dots,x_m$)
and extremely   simple deterministic actions (just
computing a polynomial). There was a hope that
Diophantine machines might be a tool for attacking
the notorious Millennium problem
$\mathrm{P \stackrel{?}{=} NP}$
about deterministic and non-deterministic
computations.

Davis provided the initial step towards the proof of his conjecture by first announcing in \cite{Davis50},
proving in his dissertation \cite{Davis50a}, and publishing in
\cite{Davis53} the following result:

\

{\bf Theorem (Davis).} \emph{
 For every listable set $\mathfrak{M}$, there exists a polynomial $Q$ with integer
 coefficients such that}
 \begin{multline}
  \langle{a_1,\dots,a_n}\rangle \in \mathfrak{M}\ \Longleftrightarrow\ 
  \exists z \forall y \le z \exists x_1,\dots,x_m\big[
  Q(a_1,\dots,a_n,x_1, \dots, x_m,y,z) = 0\big].
\label{normal}\end{multline}

\

Representations of this type became known as \emph{Davis normal forms}. They can
be considered as an improvement of Kurt G\"{o}del's technique of arithmetization.
This technique allows one to represent any listable set by an arithmetical formula containing, possibly many, universal quantifiers. Thus Davis normal form might be viewed as an elimination of all but one universal quantifiers from general arithmetical representations.
%COMMENT
{In fact, the proof in \cite{Davis53}\footnote{A footnote in \cite{Davis53} tells us that this proof was due to an unknown referee of the paper. Presumably, this was Raphael Robinson who soon afterward \cite{RRob56} quantitatively improved Davis's result by showing that one can always have $m = 4$ in \eqref{normal}.} is not an elimination but a ``compression'' of arbitrarily many universal quantifiers into a single, bounded one. The original proof given in \cite{Davis50a} and reproduced in \cite{Dav58} was quite different: Martin Davis arithmetized normal systems (introduced by E. Post \cite{Post43}) in a very economical way --- using just a single universal quantifier.}

Davis continued to work towards a proof of his conjecture together with Hilary Putnam. In his recollection \cite{Davis93a} of this collaboration during the summers 1957--1960 Davis cites a letter from Putnam: ``What I remember from that summer is not so much the mathematical details as the sheer \emph{intensity} with which we worked. I have never in my life been so absorbed in a mathematical problem, and I'm sure the same was true of you.''

 After a number of intermediate
 results,  Davis and Putnam \cite{DavisPutnam58} 
``almost proved'' the conjecture. The two missing  links were as follows.

In the first place, Davis and Putnam were forced to deal with a broader class of
\emph{exponential Diophantine representations}. They are similar to
\eqref{dio} but exponentiation is allowed (besides addition and multiplication)
for constructing a $P$.

Second, the proof given by Davis and Putnam was conditional:
they assumed that \emph{for every $k$ there exists an arithmetical progression of length $k$
consisting  of different prime numbers}.

In 1959 this hypothesis was considered plausible
but it was proved only much later, in 2004, by  Ben Green and Terence Tao \cite{GreenTao08}.
Luckily, Davis and Putnam did not have to wait that long for progress. The help came from Julia Robinson. She
 began to study Diophantine representations at the same time
 that Davis stated his conjecture. However, the direction of her original research was to show the opposite. Based on a suggestion by her professor Alfred Tarski, Robinson tried to prove that the (evidently listable) set of all powers of $2$ is \emph{not} Diophantine. After unsuccessfully attempting to prove this,
she switched to searching for Diophantine representations of this
set and,  more generally, for the set
\begin{equation}\label{abc}
 \{\langle a,b,c\rangle \mid a=b^c\}.
\end{equation}
Robinson
found a sufficient condition (later named JR by Davis) for
the existence of
such a representation.

Davis and Robinson {had} met in 1950 at
 the  International Congress of Mathematicians in
Cambridge, Massachusetts. Both of them presented their results related to Diophantine representations. Robinson recollects in \cite{Reid96}: ``I remember that he [Davis] said he didn't see how my work could help to solve Hilbert's problem, since it was just a series of examples. I said, well, I did what I could.'' In \cite{Davis93a} Davis confessed: ``It's been said that I told her that I doubted that her approach would get very far, surely one of the more foolish statements I've made in my life.''

In 1959 Davis and Putnam sent their new and still  unpublished result to Robinson. She was able to get rid of the hypothesis about arithmetical progressions of prime numbers. In retrospect, one can say that Davis, Putnam, and Robinson were under the magic of G\"odel: they were in fact trying to reproduce his technique by Diophantine equations. 
%\COMMENT
{In \cite{Davis93a} Davis writes: ``Later Yuri Matiyasevich showed that in fact any sufficiently large coprime moduli could be used so that our efforts in connection with prime factors were really unnecessary.''}

Davis and Putnam withdrew their paper and published the joint paper \cite{DPR61} with Robinson. Its main result was as follows.

\

{\bf DPR-theorem.} \emph{
Every listable set $\mathfrak{M}$  has an exponential
Diophantine representation.}

\

After that, in order to prove Davis's conjecture about \emph{arbitrary} listable
sets, it was sufficient to find a Diophantine representation for a \emph{single} special
set~\eqref{abc}. In turn, for this it was sufficient to fulfill the above-described condition
JR (found by Robinson), and Davis tried to do it.
%\COMMENT
{Davis's paper \cite{Davis68a} has the intriguing title \emph{One equation to rule them all}. In it, he proved that if a particular equation (meant in the title) has only a trivial solution, then JR is fulfilled. Robinson's attitude to this result can by seen from the following fragment of her letter to Davis which he cites in \cite{Davis93a}: ``I have enjoyed studying it, but my faith in JR still hasn't been restored. However, for the first time, I can see how it might be proved. Indeed, maybe your equation works, but it seems to need an infinite amount of good luck!''

Unfortunately, several authors later found non-trivial solutions of that equation. However this did not invalidate Davis's idea completely. It would be sufficient to establish that the set of solutions is finite. Moreover, this turned out to have some other interesting consequences.}

The first example of a Diophantine equation satisfying JR was constructed by Matiyasevich (one of the coauthors of the current paper) in \cite{Mat70b} (see also \cite{Davis73}). In February 1970, Davis got from Robinson handwritten notes that had been taken  by John MacCarthy at a lecture about this example given by Grigory Tseitin in Novosibirsk. In \cite{Davis93a}, Davis recollects: ``I was able to have the great pleasure of reconstructing the proof. But I was not satisfied until I had produced my own variant of Dr. Matiyasevich's proof and had presented it (on March 10) at a seminar at Rockefeller University at Hao Wang's invitation.'' This proof was published in \cite{Davis71}.

Having been proved, Davis's conjecture is now known under two
abbreviations:

\

{\bf DPRM-theorem ($\equiv$ MRDP-theorem).} \emph{
Every listable set   has a
Dio\-phan\-tine representation.}

\

Thus, two notions taking their origin in very different parts of science  (Diophantine equations in Number Theory and listable sets in Computer Science) determine the same class of sets of natural numbers.
 This implies that the  
concept of computability
can be defined in purely mathematical terms.

 In \cite{Davis72} Davis generalized 
 the   negative solution of Hilbert's Tenth Problem 
 in the following way: \emph{There is no algorithm for deciding whether the cardinality of the set of solutions of a given Diophantine equation belongs to a particular set unless this set or its complement is empty.}

In \cite{Davis73b} Davis transferred the so called \emph{speed-up theorem} of Manuel Blum \cite{Blum67}
to Diophantine equations.

Davis made the solution of Hilbert's Tenth Problem
available to a broad mathematical readership 
in his paper \cite{Davis73} in \emph{The American Mathematical Monthly}. A full proof can also be found
in the book \cite{Matiyasevich93a}; Davis was the editor of the English translation and wrote a \emph{Foreword} to it. Recently another book, \cite{MurtyBrandon}, with full detailed proof was published by  M. Ram Murty and Brandon Fodden.

The DPRM-theorem found numerous applications. 
Some of them  are presented in the joint paper \cite{DMR76} written by Davis, Robinson and Matiyasevich. For many other applications
of the DPRM-theorem consult the bibliography in
\cite{Matiyasevich93a} (which was extended in the
Greek translation) and in \cite{MurtyBrandon}.

The DPRM-theorem was not the final point in 
the  investigation of Hilbert's Tenth Problem.
On the contrary, the theorem opened a way
to study two natural generalizations 
of the problem:
\begin{itemize}
  \item[[Q]] \emph{is there an algorithm for solving 
  Diophantine equations in the field
 $ \mathds{Q}$ of rational numbers?}
  \item[[A]] \emph{is there an algorithm for solving 
  Diophantine equations in algebraic integers
  from a given finite extension of $ \mathds{Q}$?}
\end{itemize}
Positive solution of Hilbert's Tenth Problem 
would imply the affirmative answers to  both 
questions, but the negative solution 
has no straightforward implications to these questions.   

After numerous partial results about [A] 
(for particular fields or classes of fields)
by many authors the negative answer was 
recently announced by Peter Koymans and Carlo Pagano in \cite{koymans2024} and by Levent Alp\"oge, Manjul Bhargava,  Wei Ho, and Ari Shnidman
in \cite{alpoge2025}.

As for [Q], this question remains open in 2025.
 In \cite{Dav2010},
Davis proposed a fresh approach  to this problem by putting forth a new conjecture. It exploits not the mere existence of undecidable listable sets but the existence of so called \emph{simple sets}. The latter have been introduced by Emil Post in \cite{Post44}.

\section{Automated reasoning.}

\epigraph{We shall be concerned with the problem of testing formulas\\ in conjunctive normal form for consistency; or, dually, with\\ the problem of testing formulas in disjunctive normal form\\ for validity.}{\textit{Martin~Davis and Hilary~Putnam in \cite{DavisPutnam58a}}}

\noindent Davis was also a pioneer in a field that is called today `computational logic'. He entered this field after a quick exposure to some of the earliest computers. This journey into concrete computer programming began in 1951, when he joined a group involved in developing programs for an \textsc{ordvac} machine, supporting military efforts during the Korean War. He transitioned to the Control Systems Laboratory under Frederick Seitz after leaving his position as a Research Instructor at the University of Illinois at Urbana-Champaign. This move, coupled with a subsequent two-year Office of Naval Research grant at the Institute for Advanced Study (IAS) in Princeton, enabled him to sidestep conscription. 

Following a year-long period of intensive programming training, Davis gained sufficient confidence in his computing abilities to secure funding for a project focused on implementing a decision procedure for Presburger's integer arithmetic of addition. The procedure was implemented at the IAS, on a \textsc{johnniac} machine, during the summer of 1954.
``Its great triumph was to prove that the sum of two even numbers is even'' \cite{Davis83a}. However modest that achievement may seem today, it is often considered as the first computer-generated mathematical proof (cf. \cite{SiekmannWrightson83a}), ushering in the automated proof era. {To briefly characterize the difference in outlook between his and another mid-fifties achievement in automating deduction, namely, the Logic Theorist by Allen Newell, Herbert A. Simon (the Nobel laureate), and Cliff Shaw, in \cite{Davis83a} Davis phrases the respective slogans as ``Use mathematical logic'' and ``Simulate people''.} %\liesbeth{I believe that this is philosophically quite interesting and perhaps could be moved to the main text? I assume this comes from \cite{Davis83a}?} 

Over the years, Davis made occasional, yet significant, contributions to automated reasoning.
He also coined some of the terminology that has become standard in that field.\footnote{See, in particular, \cite{Davis63b}:
in \cite{Davis83a,Davis-Fechter-91}, Davis admitted that there he had termed ``the Herbrand Universe'' a domain---the collection of all closed terms built over a non-empty set of constants and a set of function symbols---which, he says, should more correctly be called ``the Skolem Universe.''}

The report \cite{DavisPutnam58a}, written jointly by Martin Davis and Hilary Putnam in 1958, evolved into articles (\cite{DavisPutnam60}, \cite{Chinlund-Davis-etAl64}, etc.) that furthered its broad influence on later research. That report arose from a project that Davis and Putnam, initially willing to tackle proof automation in an all-encompassing framework (undecidable, albeit semi-decidable\footnote{Given a formal statement, a decision algorithm should tell whether or not it is provable as a theorem; a semi-decision procedure may, when confronted with a false or unprovable statement, run forever.}), had proposed to the National Security Agency. ``When we talked about our proposal---Davis recalls in \cite{Davis20}---, they made it clear that they had no interest in proof procedures for first order logic, but they were interested in SAT.\footnote{Authors' note. SAT (short for `satisfiability') is the problem of whether there is a way to assign truth values to the variables in a formula of propositional logic that makes the formula evaluate to `true'. Our discourse will soon shift to a more general notion of satisfiability, which refers to formulas of first-order logic.} They warned me that it is a difficult problem and doubted that we could make much headway in one summer. However if we were willing to work only on SAT, they were prepared to fund our proposal.'' As an outcome, the afore-cited 1958 deliverable approaches propositional calculus from a broad angle. Its first part discusses the advantage of bringing formulas to some standard format; it notes that not all such forms have the same properties and argues that conjunctions of disjunctions of propositional variables and negations of such  (\emph{CNF} formulae, for short) are a convenient format to be treated by satisfiability testing methods. Embedding a CNF-tester in a proof procedure for full quantification theory later enabled the approach by Davis and Putnam to outperform competitors of the time (see \cite{DavisPutnam60}), particularly the theorem-proving programs developed by Paul C. Gilmore \cite{Gil60}, Hao Wang \cite{Wang60}, and Dag Prawitz et al. \cite{PPV60}.

Two CNF-satisfiability testers (see \cite{DavisPutnam60} and \cite{DavisLogemannLoveland62a}), whose acronyms are DP and DPLL after the names Davis--Putnam--Logemann--Loveland, evolved from the said seminal study \cite{DavisPutnam58a}.
Interestingly, the DPLL decision procedure remains at the core of fast Boolean satisfiability solvers decades later (see \cite{LSS16} and \cite{BHMW21}). Furthermore, the problem of recognizing CNF-satisfiability has become the standard for representing
the difficulty of solving by a feasible algorithm a class of hard combinatorial %(NP-hard)
problems---a millennium-prize challenge.\footnote{See \url{https://www.claymath.org/millennium/p-vs-np/}} 

To briefly describe the functioning of DPLL, let us assume that the CNF formula to be analyzed is input as a set $\mathcal{S}_0$ whose members are sets of nonnull integers: positive integers encode propositional variables and negative integers represent their opposites. The set $\mathcal{S}_0$, as well as its members, henceforth named \emph{clauses}, are finite; the members of the clauses are called \emph{literals}. Clauses represent the disjunctions that form the given formula. As extreme cases, the empty set of clauses represents the truth value `true' whereas any set to which the empty clause belongs represents `false'. 

\emph{Simplifying} a clause set $\mathcal{S}$ with respect to some number $\ell$ means removing all clauses to which $\ell$ belongs from $\mathcal{S}$, while removing its opposite, $-\ell$, from the remaining clauses. Performing such a simplification means assigning the value `true' to the literal represented by $\ell$; thus, all clauses to which $\ell$ belongs become true, while $-\ell$ can no longer help any other clause to become true.

DPLL starts by putting $\mathcal{S}=\mathcal{S}_0$ and repeatedly simplifies $\mathcal{S}$ until it becomes empty, in which case the initial $\mathcal{S}_0$ is declared satisfiable and the algorithm terminates. Each simplification step extends a truth-value assignment intended to make $\mathcal{S}_0$ true: When simplifying w.r.t. a literal $\ell$, it is choosing to make $\ell$ true and prevents $-\ell$ from being chosen later. Therefore, DPLL does not just establish satisfiability but also produces a model for $\mathcal{S}_0$---if any.

The case when $\mathcal{S}$ becomes false (because one of its clauses empties) is asymmetric: the algorithm backtracks to a \emph{choice point}---specifically, a situation where the literal $\ell$ that triggered a simplification was preferred over its unattempted opposite, $-\ell$. At this point, the alternative simplification is triggered. When no choice points remain and $\mathcal{S}$ has not emptied, $\mathcal{S}_0$ is declared unsatisfiable.

Let us go into more detail. DPLL orchestrates the following three rules, giving priority to the first two over the third.\footnote{Additional heuristics were present in the original implementation.}
\begin{itemize}
\item[-]\emph{Unit rule:}\ After finding a singleton clause $\{\ell\}$ in $\mathcal{S}$, simplify $\mathcal{S}$ w.r.t. $\ell$.
\item[-]\emph{Pure-literal rule:}\ After finding some literal $\ell$ belonging to a clause of $\mathcal{S}$ such that $-\ell$ does not belong to any of its clauses, simplify $\mathcal{S}$ w.r.t. $\ell$.
\item[-]\emph{Splitting rule:}\ Pick a literal $\ell$ such that both $\ell$ and $-\ell$ belong to clauses of $\mathcal{S}$; set a choice point here. Tentatively simplify $\mathcal{S}$ with respect to $\ell$ and continue. If subsequent simplifications do not lead to satisfying $\mathcal{S}$, undo the previous simplification and, instead, simplify $\mathcal{S}$ w.r.t. $-\ell$\/.
\end{itemize}

It deserves being mentioned that the original DP algorithm had, in place of the splitting rule, a rule based on the following remark (which somehow anticipates the Resolution principle --- see below):

``{\emph{Rule for Eliminating Atomic Formulas.}\ Let the given formula $F$ be put into the form $(A\ \vee\ p)\ \&\ (B\ \vee\ \bar{p})\ \&\ R$ where $A,B$, and $R$ are free of $p$. (This can be done simply by grouping together the clauses containing $p$ and `factoring out' occurrences of $p$ to obtain $A$\/, grouping the clauses containing $\bar{p}$ and `factoring out' $\bar{p}$ to obtain $B$\/, and grouping the remaining clauses to obtain $R$.\/) Then $F$ is inconsistent if and only if $(A\ \vee\ B)\ \&\ R$ is inconsistent.}'' (Martin~Davis and Hilary~Putnam in \cite{DavisPutnam60}.

 It is reported in \cite{DavisLogemannLoveland61} that after programming DP using this rule, it was decided to replace it by the splitting rule: the two are interchangeable, but the former proved ``prohibitive in a computer if one's fast access storage is limited''.

\medskip

With regard to first-order logic, Davis devised a method named \emph{Linked Conjunct} to automatically generate a proof of a sentence by seeking a counterexample to its negation. That method was implemented in a LISP-based theorem prover at Bell Labs --- see \cite{Davis63b,Chinlund-Davis-etAl64}. It borrowed from the above-mentioned Prawitz prover the idea of `mating'---through a suitable algorithm based on the so-called \emph{unification}\footnote{A unification algorithm's task is to find a substitution of individual variables with terms that makes two given terms or atomic formulas syntactically identical. If no such substitution exists, the algorithm simply fails; otherwise, it produces one of maximum generality. Mechanisms for this purpose were independently devised around 1960 by various authors (including Davis's team at Bell Labs), unaware that Jacques Herbrand had already sketched a unification algorithm in his Ph.D. thesis in 1930. As pointed out in Davis's introductory paper \cite{davis+emil+1994} to \emph{The Collected Works of Post}, also Post had a unification algorithm but it was unpublished.}---instances of atomic formulae bearing complementary signs in different conjuncts, i.e., disjunctive clauses. Even though the Linked Conjunct method became less popular after the more influential Resolution principle by John Alan Robinson gained prominence, some improved variants of the latter can be best explained from the viewpoint of the former method (cf. \cite{Yarmush76} and \cite{Omo82}), as we will now highlight.

Automatic theorem-proving procedures usually presuppose that the axioms of a theory together with the negation of a theorem's claim are converted into an \emph{unsatisfiable} conjunctive form;\footnote{We use the term `conjunctive form' instead of `CNF' in the present context to emphasize that we are referring to first-order logic. Here, within formulas, `atoms' of the form $R(t_0,t_1,\dots,t_n)$---where $R$ stands for an $(n+1)$-argument predicate symbol and each $t_i$ represents a term---play a role analogous to the true-false variables of propositional logic.} their main task is to find a \emph{refutation} of the set of (disjunctive) clauses representing this conjunctive form, where no quantifier appears, but every individual variable is tacitly regarded as universally quantified. There are several notions of refutation to choose from. For example, the individual steps that form a refutation may generate clauses or conjunctions thereof; correspondingly, the final step of a refutation leads to the empty clause or to a \emph{truth-functionally} unsatisfiable set of clauses. In the latter case, one may insist---as Davis did, with no loss of generality---that every literal occurs both affirmed and negated within the final step of a refutation (this is the `linkedness' property referred to in the name of Davis's method).

When the steps of a refutation generate clauses (as is the case with strictly resolution-based methods), each application of an inference rule removes one or more occurrences of literals from a set of unifiable occurrences. The loss of information due to this deletion is one of the causes of the creation of an intractable number of useless clauses during the search for a refutation. This `combinatorial explosion' turns out to be the main drawback of resolution-based methods for automatic theorem proving.

In the Linked Conjunct method, on the other hand, the detection of a pair of complementarily unifiable literals does not lead to literal deletion,  but rather to the `mating' of two literals with one another. Thus, in a more precise description than the one given above, a single step of a refutation manipulates a set of clauses with a superimposed mating relationship holding between pairs of occurrences of literals (not to mention a binding of terms to individual variables). As a support for such manipulations, graph-like structures can conveniently represent the set of clauses. Such graphs, which the Linked Conjunct method handles explicitly, serve mainly as a conceptual tool for demonstrating the completeness of sophisticated refinements of Robinson's resolution method.

\medskip 

Over the decades, there has been a noticeable shift in expectations regarding proof-assistant systems. These systems have evolved from stand-alone theorem provers into proof-checking frameworks that range from highly interactive reasoning assistants to simple proof-script verifiers. Davis played a role in catalyzing this shift in perspective.

In \cite{davis-schwartz79a}, Davis and Jacob T. Schwartz advocated for incorporating \emph{metamathematical extensibility} mechanisms into mature proof-verification technology, enabling proof-checkers to undergo substantial enhancements without compromising their reliability.

Around 1980, Davis spent time at John McCarthy's Stanford Artificial Intelligence Laboratory (SAIL). 
In the lab, he also worked with the FOL (`First Order Logic') proof checker developed by Richard Weyh\-rauch. Referring to that experience, he said that while he enjoyed the quality of interaction with the automated assistant, he also felt irritated by  the need to make many painstaking tiny steps to justify inferences that were quite obvious. As a result, by reusing the LISP source code for the linked-conjunct theorem prover, a Stanford undergraduate successfully implemented an `obvious' facility as an add-on to FOL, see \cite{Davis81obviouslogical}. In addition, \cite{Davis80b} was also written during his time in what Davis found to be a very stimulating environment.

\section{Nonstandard analysis}\label{sec:nonstandardAnalysis}
\epigraph{Perhaps indeed, enthusiasm for nonstandard methods is not unrelated to the well-known pleasures of the illicit.}{Martin~Davis in \cite{Davis77}}
    
\noindent Davis recalled in \cite{Dav16}: ``As an undergraduate I had tried briefly to rehabilitate Leibniz's use of infinitesimal quantities as a foundation for calculus [$\dots$] It was therefore with great excitement and pleasure that I heard Abraham Robinson's address before the Association for Symbolic Logic towards the end of 1961 in which he provided an elegant solution to this problem using techniques that he dubbed \emph{nonstandard analysis}.'' 

After Davis and Reuben Hersh wrote, on this subject, the popular article \cite{DavisHersh72} for The Scientific American magazine, Davis published the monograph \cite{Davis77} which appeared in print at the same time as two other major books on infinitesimal calculus
(cf. \cite{blass1978}).

In the introduction of \cite{Davis77}, Davis stated: ``The subject can only be claimed to be of importance insofar as it leads to simpler, more accessible expositions, or (more important) to mathematical discoveries. As to the first, the reader must be the judge. The best evidence for the second is the Bernstein--Robinson theory [$\dots$], which settled a question that had remained open for many years.'' The 
book in fact culminates in the proofs, by nonstandard methods, of important theorems about Hilbert spaces, including the Bernstein--Robinson theorem.

The way in which Davis conceived the initial part of this monograph offers a brilliant example of information hiding as a guiding principle for the design of widely applicable constructions and methods of proof. As emphasized in \cite{CantoneOP16}, the concept of withholding details, as stated here, is treated in another 1977 writing: the previously cited \emph{Metamathematical extensibility for theorem provers and proof-checkers} \cite{davis-schwartz79a}, which was more related to proof technology than to the foundations of analysis.

The first chapter of \cite{Davis77} concentrates, in fact, on how to enlarge a standard `universe' (this word has a specific technical sense) into a nonstandard one. While taking stock at the end of this chapter, Davis remarks that much of the machinery developed up to that point will play no role in the rest of the book; then, in recapitulating which key issues the reader should bear in mind, he stresses three main tools of nonstandard analysis: \emph{transfer principle}, \emph{concurrence}, and \emph{internality}. Concerning the elaborate construction in which he had %been entangled 
focused on for over forty pages---the so-called \emph{ultra-power} construction---, he concludes: ``The reader who remembers these key points will do well in what follows. In particular, it is now quite all right to entirely forget how the nonstandard universe was defined and to banish ultrafilters from our consciousness''.

\section{History and philosophy of computing.}\label{Sec:HistoryAndPhilosophy} Davis not only lived through a large part of the early history of
computing and computer science, having studied with Post and Church, he also cared deeply about preserving that
history and making that history accessible to students with different levels of mathematical maturity. The volume \emph{The Undecidable}, mentioned above, contained two unpublished
papers, namely, G\"odel's Princeton Lectures ``On undecidable propositions of formal mathematical systems'' and Post's ``Account of an anticipation'', as well as 
fifteen other published, pioneering papers in
computability theory by Alonzo Church, Stephen C. Kleene, J. Barkley Rosser, and Alan
M. Turing, together with other papers by G\"odel and Post. Each of the seventeen texts
was accompanied by a short introductory note, contextualizing it in the landscape of
early computability theory. These introductory notes succeed in illuminating this
intellectual landscape for a broad audience. As mentioned above, Davis later edited
Post's \emph{Collected Works} for which he wrote an extended biographical introduction --- a
testimony to Davis's admiration and respect for his teacher.

Davis's contributions to the history of computing were not restricted to ``just'' editing important documents but included writing scholarly papers on topics that reflected his deep interest in logical foundations and the emergence of computer science. The resulting sequence of papers includes, among others, ``Why G\"odel didn't
have Church's thesis'' \cite{Davis82}. At the core of this study are basic questions related to the early
history of the so-called Church--Turing thesis.\footnote{We say ``so-called'' since, historically, there is no such thing as the Church--Turing thesis. Cp. Note \ref{footnote:ChurchThesis}.
In his \cite{Kleene1952}, Kleene had devoted two sections for the separate insightful discussion of Church's, respectively Turing's thesis.
} As Davis explains, it was prompted by Kleene's  work \cite{Kleene81} on
the origins of recursive functions and the more philosophical work Webb \cite{Webb80}
about the consequences of the ``Church--Turing thesis'' for issues related to mechanism and
mentalism. Using historical documents and personal letters, Davis explored  why
G\"odel was hesitant to support a thesis similar to Church's, while Church was ``willing
to go ahead''.\footnote{Sieg in \cite{Sieg97}, having access to additional documents, showed that Church was initially  also quite hesitant.} Moreover, Davis is careful to distinguish between  Church's, G\"odel's, Kleene's, Post's and Turing's theses and provides a much needed historically and conceptually diversified view on the so-called ``Church-Turing thesis''.  By reconstructing the formulation of these different theses, Davis also provides an account of how Post anticipated the results of Church, G\"odel and Turing in the early 1920s.\footnote{The reader might wish to compare this paper with Gandy's \textit{The confluence of ideas in 1936} \cite{gandy+confluence+1988}. Davis and Sieg were alluding to the title of that paper when emphasizing in their \cite{DavSieg16} the remarkable conceptual confluence of Post's and Turing's work.}  This paper remains a classic for anyone working on the early
history of the ``Church--Turing thesis''.

While Davis worked on the history of ``pure'' computability theory, he also wrote
extensively on the significance of logic for the emergence of computer science and
the modern computer. The earliest publication on those topics is his 1987-paper
``Mathematical logic and the origin of modern computers''. Perhaps, best-known is his
book \emph{The Universal Computer: The Road from Leibniz to Turing} that was published in 2000. A new version was published in 2012, the Turing centenary. There, Davis inscribes the history of the modern computer in a much longer history of logic and mathematics going back to Leibniz' notions of a \textit{characteristica universalis} and a \emph{calculus ratiocinator}. He develops from there a narrative of the history of logic and mathematics that leads to the work of G\"odel and Turing and shows how logic and computation are ``\textit{two sides of the same coin}'' \cite[p.~200]{Davis00}. He then argues that this historical path from Leibniz to Turing laid the foundation for the modern computer. The topic of the influence of logic on computing, in particular the significance of Turing's work in this context, has been a difficult and indeed controversial one amongst computer scientists and historians alike; Davis' 2012 edition too got caught up in that debate.\footnote{The significance of Turing's 1936 paper for the actual design and construction of the first computers is debatable; there is mostly consensus among historians that Turing is, at best, one of many who contributed to that development. As was argued in \cite{DeMol2018}, the introduction of the Turing machine concept as a model for modern computers only came in the early 1950s. --- The close ties between logic and computer science were loosened by emphasizing other important contributions in \cite{Mahoney1988}. Davis' depiction of those ties was criticized, for example, in \cite{Daylight2014}.} 

But his work should also be seen in its proper context: at the time Davis started to write about these topics, it was one of the \textit{only} historical works available on the topic. %Davis put great efforts in his arguments and the historical evidence on which they were based. 
Davis's work here was heavily influenced by 
his own experience as a logician who
became involved with programming early on, and whose work on automatic theorem
proving was later seen as fundamental in computer science. This work  provides an exemplar of the perspective of a computer scientist who was trained as a logician and discovered that
his work could be recast as belonging to a subject that did not exist at the
time of his training.

We already argued, in \Sec{Sec:computabilityAndUnsolvability}, that Davis was strongly convinced of the correctness of the Turing--Post analysis of computability. In fact, Davis was explicit that he was a mechanist and so convinced that human brains cannot surpass incompleteness \cite{Jac08,Davis12}. This methodological and philosophical stance made him an outspoken critic of ``hypercomputability'' as formulated, among others, by Copeland \cite{Copeland00} and Siegelmann \cite{Siegelmann99}. They considered versions of the Church--Turing thesis as false, because they constructed some ``physically realizable'' models that can ``compute'' more than Turing machines can. Davis criticized these works on multiple occasions through conceptual and historical arguments but avoiding dogmatism.\footnote{See for instance \cite{Davis04,Davis06a}} Amongst others, he argued that such a ``result'' can be obtained only when finiteness conditions are violated that are essential in Turing's analysis. Being clear about this restriction, Davis is open to the possibility of an expanded, different notion of ``computability''. In one of his pieces on hypercomputation he writes: ``Can we really utterly exclude the possibility of some new development leading to a physical realization of an uncomputable quantity? Of course not. [\dots]'' \cite[p.206]{Davis04}. When questions such as ``Is the mind a computer?'' or ``Is human insight algorithmic?'' were raised, Davis took a similarly open position. For instance, Roger Penrose argued in two books \cite{Penrose89,Penrose94} that G\"odel's incompleteness theorem implies that the human mind cannot be a computer. Davis rejected Penrose's claims with clear technical arguments. He did this in two short papers, \cite{Davis90} and \cite{Davis93c}. In the first, he concludes: ``There is certainly room for disagreement about whether the processes by which mathematical (or physical) theories are developed and accepted are algorithmic. But G\"odel's theorem has nothing decisive to contribute to the discussion.'' These papers reveal Davis as a writer who could not stand what he considered to be false or badly argued claims but also as a thinker who is careful when speaking about questions that cannot be solved by proof alone. In such contexts, he often referred to an empiricist or inductive position. That position is elaborated in what is perhaps Davis' most explicitly philosophical paper. There it becomes clear how Davis' anti-dogmatism is intertwined with this more empirical view on mathematical practice, for instance, when he writes \cite{Davis12}: 
\begin{quote}
We explore simple austere worlds that differ from the one we inhabit both by their stark simplicity and by their openness to the infinite. It is simply an empirical fact that we are able to obtain apparently reliable and objective information about such worlds. And, because of this, any illusion that this knowledge is certain must be abandoned. [...] If presented with a proof that Peano Arithmetic is inconsistent or even that some huge natural number is not the sum of four squares, I would be very very skeptical. But I will not say that I know that such a proof must be wrong.
\end{quote}
Davis' anti-dogmatic empiricism is also what drove his more ``unconventional'' position with respect to the P vs. NP problem. He repeatedly stated that, from his view, there is no \textit{convincing} heuristic evidence that $P \neq NP$.\footnote{See for instance \cite{Jac08}.}

The \emph{The Universal Computer} treats the history of computability theory no longer narrowly as being equivalent to that of recursion theory. On the contrary, the subject is connected to penetrating questions and claims of Leibniz, to foundational problems in 19$^{\rm th}$ and early
20$^{\rm th}$ century mathematics, as well as to contemporary issues pertaining to
computer technology, cognitive psychology and philosophy of mind. Davis's goal was not only to provide a clear, non-technical exposition of the development of
computability theory and its central results, but also to describe their broad impact on
contemporary thought. More importantly, perhaps, Davis wanted to bring out how
logic, deeply embedded in philosophy, articulated foundational problems and
addressed them from a genuinely revolutionary perspective. At the end of the book's
Introduction, Davis wrote: ``Nowadays, as computer technology advances with such
breathtaking rapidity, as we admire the truly remarkable accomplishments of the
engineers, it is all too easy to overlook the logicians whose ideas made it all possible.
This book tells their story.'' For us it is clear that Davis became through his work
an integral part of ``their story'' --- deeply rooted in the intersection of philosophy,
mathematics, and computer science.

\bibliographystyle{plainurl}

\bibliography{main}  

\end{document}